\documentclass[%
 reprint,
superscriptaddress, 
 amsmath,amssymb,
aps, 
pra, 
]{revtex4-1}

\usepackage{graphicx}
\usepackage{dcolumn}
\usepackage{longtable} 
\usepackage{multirow}
\usepackage{hhline} 
\usepackage{rotating}
\usepackage{bm}
\usepackage{natbib}
\usepackage{gensymb} 

\begin{document}

\title{Tailored nanodiamonds for hyperpolarized $^{13}$C MRI}


\author{T. Boele}
\affiliation{ARC Centre of Excellence for Engineered Quantum Systems, School of Physics, University of Sydney, Sydney, NSW 2006, Australia}
\author{D. E. J. Waddington}
\affiliation{ARC Centre of Excellence for Engineered Quantum Systems, School of Physics, University of Sydney, Sydney, NSW 2006, Australia}
\affiliation{ACRF Image X Institute, Faculty of Medicine and Health, University of Sydney, Sydney, NSW 2006, Australia}
\author{T. Gaebel}
\affiliation{ARC Centre of Excellence for Engineered Quantum Systems, School of Physics, University of Sydney, Sydney, NSW 2006, Australia}
\author{E. Rej}
\affiliation{ARC Centre of Excellence for Engineered Quantum Systems, School of Physics, University of Sydney, Sydney, NSW 2006, Australia}
\affiliation{Department of Physics, California Institute of Technology, Pasadena, CA 91125, USA}
\author{A.~Hasija}
\affiliation{Department of Molecular Sciences, Macquarie University, Sydney, NSW 2109, Australia}
\author{L. J. Brown}
\affiliation{Department of Molecular Sciences, Macquarie University, Sydney, NSW 2109, Australia}
\author{D. R. McCamey}
\affiliation{ARC Centre of Excellence for Exciton Science, School of Physics, University of New South Wales, Sydney, NSW 2052, Australia}
\author{D. J. Reilly}
\affiliation{ARC Centre of Excellence for Engineered Quantum Systems, School of Physics, University of Sydney, Sydney, NSW 2006, Australia}
\affiliation{Microsoft Quantum Sydney, University of Sydney, Sydney, NSW 2006, Australia}


\begin{abstract}
Nanodiamond is poised to become an attractive material for hyperpolarized $^{13}$C MRI if large nuclear polarizations can be achieved without the accompanying rapid spin-relaxation driven by paramagnetic species. Here we report enhanced and long-lived $^{13}$C polarization in synthetic nanodiamonds tailored by acid-cleaning and air-oxidation protocols. Our results separate the contributions of different paramagnetic species on the polarization behavior, identifying the importance of substitutional nitrogen defect centers in the nanodiamond core. These results are likely of use in the development of nanodiamond-based imaging agents with size distributions of relevance for examining biological processes. 
\end{abstract}

\maketitle

\section{\label{sec:level1}INTRODUCTION}
Hyperpolarized $^{13}$C magnetic resonance imaging (MRI) leverages a $>$10,000-fold enhancement in $^{13}$C polarization achieved via dynamic nuclear polarization (DNP), a process in which spin polarization is transferred from electron spins to $^{13}$C nuclei \cite{Ardenkjaer-Larsen2003}. Hyperpolarized modalities have recently enabled metabolic imaging of the heart as well as tumors of the brain and prostate \cite{Cunningham2016,Bastiaansen2016,Nelson2013,Miloushev2018}. Despite the new diagnostic methods these imaging techniques offer, they are limited by the short spin-lattice relaxation times ($T_1$) of the liquid-state metabolic compounds that restrict the lifetime of the hyperpolarized signal \cite{Keshari2014,Chattergoon2013}.

Solid-state nanoparticles offer an imaging modality comparatively unlimited by $T_1$. Two promising candidates are silicon and diamond, which have dilute spin systems of $^{29}$Si and $^{13}$C at 4.7 and 1.1\% natural abundance respectively. Both silicon and diamond nanoparticles have been investigated and reported to maintain $T_1$ relaxation times that approach the hours-long $T_1$ times  of their bulk counterparts \cite{Atkins2013,Rej2015,Shulman1956,Hoch1988}. Silicon and diamond nanoparticles can also be hyperpolarized via DNP using their endogenous paramagnetic defects as a source of free electrons \cite{Rej2015,Dutta2014,Casabianca2011,Bretschneider2016,Cassidy2013A,Dementyev2008,Yoon2019} and still have sufficiently long spin-spin relaxation times ($T_2$) to allow for MRI with useful spatial resolution \cite{Cassidy2013B,Whiting2016,Kwiatkowski2018,Waddington2019}. 

Paramagnetic defects that drive DNP however, also lead to spin relaxation and a shorting of $T_1$. Finding nanoparticles of biologically-relevant size that combine long $T_1$ with the ability to create high $^{13}$C polarization has proven to be challenging. With the correct balance of paramagnetic defects to fuel DNP without overly accelerating relaxation, hyperpolarized nanodiamond holds the promise of a biocompatible \cite{Moore2016,Lee2017} MRI imaging agent possessing the advantages over silicon of optical trackability \cite{Chow2011,Xia2018,Xi2014,Gu2018} and a readily-functionalized, non-oxidizing surface \cite{Mochalin2011,Nunn2017}. Although previous works have investigated $^{29}$Si defects and DNP \cite{Kwiatkowski2018B}, as well as nitrogen-vacancy defects in diamonds, including recent work hyperpolarizing diamond powders via optical methods \cite{Doherty2013,Ajoy2018}, little has been done to investigate the optimum defect concentration for nanodiamond DNP. 

Here we report experiments investigating the effect of paramagnetic defect concentration and defect type upon DNP performance in nanodiamond. Surface modification provides a practical means of altering the paramagnetic makeup of nanodiamonds, taking advantage of the large surface-to-volume ratio of nanoparticles. This method is shown to be effective at altering the defect composition, as it selectively removes those defect types associated with the nanodiamond surface, leaving defects associated with the core of the nanodiamond untouched. Furthermore, since the effects of surface modification upon the nuclear magnetic resonance (NMR) properties of nanodiamonds are well understood, aspects that relate specifically to DNP performance can be simply separated \cite{Panich2017}.

We begin by demonstrating that DNP is effective for hyperpolarizing a wide range of nanodiamonds of different sizes, defect compositions and concentrations. We find that using acid-cleaning and air-oxidation to purify the surface affects the DNP performance in a way that depends on the size of the nanodiamond. Surface modification increases the maximum $^{13}$C polarization achieved in nanodiamonds hundreds of nanometers in diameter and decreases the $^{13}$C polarization achieved in nanodiamonds tens of nanometers in diameter. By examining these results we identify the importance of substitutional nitrogen P1 centers distributed throughout the nanodiamond core. Optimal DNP performance is achieved for nanodiamonds with a core of P1 centers and a surface cleaned of excess paramagnetic defects. Since the effectiveness of a particular defect for performing DNP is largely related to the properties of its paramagnetic spin, we also examine electron relaxation times for different defects and conclude that the extended spin-lattice relaxation time of the P1 center is largely responsible for its superior contribution to effective DNP of nanodiamond. Taken collectively, these results provide a guide to the creation of the optimal nanodiamond for use in hyperpolarized $^{13}$C MRI.

\section{\label{sec:level1}RESULTS AND DISCUSSION} 

\subsection{\label{sec:level2}Endogenous defect composition}

\begin{figure*}[htb]
 \centering
   \includegraphics[width=1\textwidth]{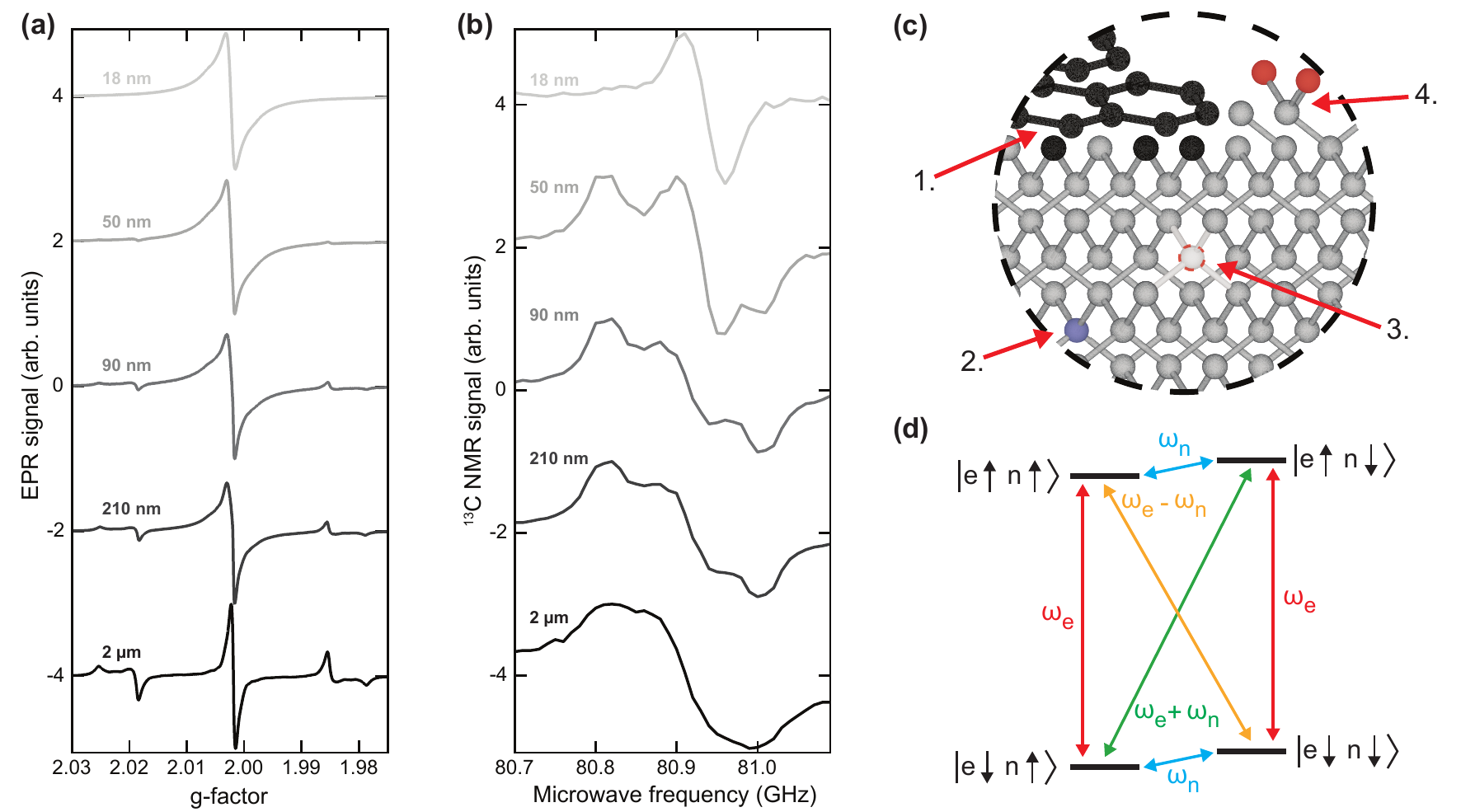}
 \caption{Hyperpolarizing nanodiamond via endogenous paramagnetic defects. \textbf{(a)} EPR spectra for a range of monocrystalline synthetic nanodiamond samples in increasing size from top to bottom. Spectra are normalized and offset for clarity. \textbf{(b)} DNP spectra corresponding to the nanodiamond samples in (a) illustrating the different DNP behavior exhibited by nanodiamonds of different size and composition. \textbf{(c)} Cross-section of a portion of a nanodiamond lattice illustrating the nanodiamond structure. The core of the nanodiamond is made up of carbon atoms in a tetrahedral arrangement (shown in grey) of sp$^{3}$-hybridized atoms. The nanodiamond surface is covered with a layer of functional groups or graphitic, hexagonal sheets of sp$^{2}$-hybridized carbon (shown in black) that forms chains and patches. 1. Graphite on the surface. 2. P1 center (showing substitutional nitrogen in blue). 3. Vacancy. 4. Carboxylic COOH group (showing oxygen atoms in red). During acid cleaning the graphite is etched away and the surface is oxidized. The end result is a surface terminated with COOH groups. \textbf{(d)} Solid effect energy level diagram. The DNP spectra in (b) are well explained by convolving the absorption EPR spectra and a solid effect enhancement spectrum with a maximum at $\omega{}_e-\omega{}_n$ and a minimum at $\omega{}_e+\omega{}_n$.}
 \label{FIG1}
\end{figure*}

As nuclear $T_1$ relaxation and DNP behavior depend on the paramagnetic defects in nanodiamond we begin our results with continuous wave electron paramagnetic resonance (CW EPR) characterization measurements.  A representative set of EPR and DNP spectra is shown in Fig. \ref{FIG1}  to illustrate the variation exhibited over the range of nanodiamond sizes investigated. Data taken at room-temperature using X-band EPR are shown in Fig. \ref{FIG1}a for monocrystalline synthetic diamond powders between 18 nm and 2 $\mu$m in size, revealing a transition from a spectrum with a single dominant spin-$1/2$ EPR transition to a more complicated spectrum with multiple contributions.

The EPR spectra can be well modeled by a three-spin model made up of a broad spin-$1/2$ component, a narrow spin-$1/2$ component, and a P1-center component. We attribute the broad component to spins associated with dangling bonds on the nanodiamond surface and the narrow component to dislocations and vacancies in the diamond lattice. The P1-center component accounts for single substitutional nitrogen atoms in the diamond. For P1 centers the hyperfine coupling between the extra valence electron of the nitrogen atom and the spin-1 $^{14}$N nucleus creates the distinctive antisymmetric features either side of the central resonance that become more pronounced for larger diamond particles. The cartoon cross-section Fig. \ref{FIG1}c shows the type of features that give rise to the different contributions in the EPR spectrum.

To illustrate the changes in DNP behavior with varying diamond particle size and paramagnetic defect composition, Fig. \ref{FIG1}b plots DNP spectra acquired at 2.89 T, 4.5 K for the diamond samples with EPR spectra shown in Fig. \ref{FIG1}a. For the smallest nanodiamond sample a typical solid effect profile is observed, with lobes of positive and negative enhancement \cite{Wenckebach2008}. These lobes correspond to the sum and difference of the electron and $^{13}$C nuclear resonance frequencies ($\omega{}_e$ and $\omega{}_n$ respectively) providing positive enhancement of nuclear polarization via the $\omega{}_e-\omega{}_n$ transition and negative enhancement via the $\omega{}_e+\omega{}_n$ transition, as shown in Fig. \ref{FIG1}d. Increasing the size of the nanodiamond leads to additional features appearing in the DNP spectrum. These are associated with the increasing contribution from the P1-center electrons that split the coupled electron-nuclear energy levels (shown in Fig. \ref{FIG1}d) and give rise to additional frequencies that provide enhancement of $^{13}$C polarization. 

To explain the shape of the DNP spectrum we consider the EPR linewidths at the magnetic field strength in the polarizer rather than at X-band ($\approx$0.35 T). At 2.89 T the expected EPR inhomogeneous linewidth of the P1 centers in our nanodiamond samples is 17 MHz, compared to 56 MHz for the broad spin-$1/2$ component and 31 MHz for the $^{13}$C NMR frequency \cite{Yavkin2015}.  As the expected width of the total EPR spectrum at 2.89 T is greater than $\omega{}_n$, DNP can simultaneously drive $\omega{}_e+\omega{}_n$ and $\omega{}_e-\omega{}_n$ transitions and the net result is the difference between the competing pathways. This is the differential solid effect and combined with the multi-component EPR spectrum of our nanodiamond it accounts qualitatively for the shape of the DNP spectrum \cite{Wenckebach2016}.

\begin{figure}[htb]
 \centering
   \includegraphics{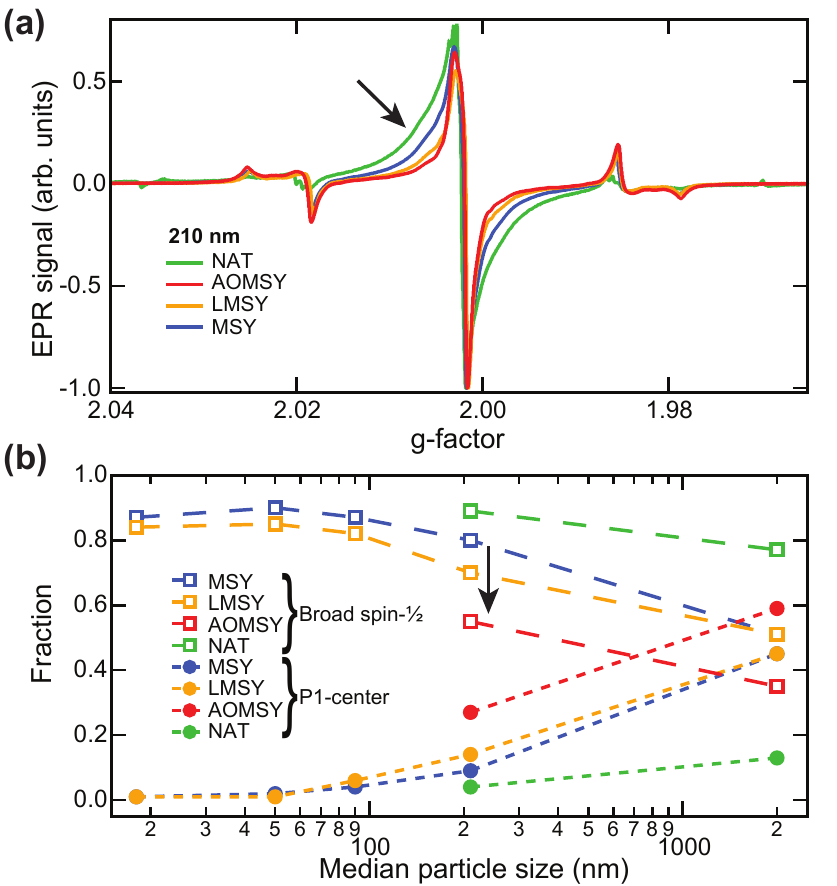}
 \caption{Paramagnetic defect characterization and the effect of surface treatments. \textbf{(a)} EPR spectra for 210 nm natural (green), air-oxidized (red), acid-cleaned (yellow), and untreated monocrystalline (blue) nanodiamond samples. We draw attention to the portion of the EPR spectrum that shows a marked decrease following surface treatment. Spectra are normalized to aid comparison of their shape. \textbf{(b)} Fraction of total spins made up by broad spin-$1/2$ and P1-center spins as a function of particle size. Data points shown are extracted from the weights found in the fitting process described in the methods section. Acid-cleaning and air-oxidization processes remove paramagnetic defects from the diamond surface, reducing the fraction of spins that contribute to the broad spin-$1/2$ component in the EPR spectrum. As particle size increases the fraction of P1-center spins (solid circles) increases for all diamond types. Error bars from fitting error are smaller than data points.}
 \label{FIG2}
\end{figure}

To further investigate how defect composition affects nanodiamond DNP we have applied two different oxidation techniques to alter the nanodiamond surface. We have performed anaerobic oxidation by applying several acid purification cycles and aerobic oxidation by heating nanodiamonds in a furnace in air. By removing graphitic carbon and oxidizing the nanodiamond surface, acid-cleaning and air-oxidation techniques change the paramagnetic composition of the diamond, diminishing the contribution of the broad spin-$1/2$ component to the EPR spectra, as shown in Fig. \ref{FIG2}. This effect is highlighted in Fig. \ref{FIG2}a, which compares the EPR spectra of surface treated synthetic, untreated synthetic, and natural 210 nm nanodiamond samples. The comparison clearly shows that the central EPR transition was narrowed by acid cleaning and narrowed further by air oxidation, whilst natural nanodiamonds have a greater proportion of their electrons made up by the broad component. Fits to these results, presented in Table \ref{TAB1} and plotted in Fig. \ref{FIG2}b, confirm that the fraction of electron spins attributed to the broad spin-$1/2$ component of the EPR spectrum decreased following surface treatment and was largest for natural nanodiamond samples. A contingent effect was that the fraction of P1-center spins increased as the broad spin-$1/2$ fraction decreased. Combined with the range of defect concentrations exhibited by nanodiamonds of different size shown in Table \ref{TAB1} this control of defect composition provides a means of investigating the optimum defect type and concentration for DNP.

\begin{table*}[htb]
\begin{ruledtabular}
\begin{tabular}{ l l l l l l l }

\multirow{2}{*}{Diamond} & Size range & Median & Defect concentration & \multicolumn{3}{c}{Fraction of spins}\\

 & ($\mu$m) & size &  ($10^{19}$ spins/g) & Narrow & Broad & P1 center \\
\hline
\multirow{5}{*}{Monocrystalline synthetic (MSY)} & 0-0.03 & 18 nm & 3.2 $\pm$ 0.6 & 0.13 & 0.87 & 0.01 \\
& 0-0.1 & 50 nm & 5.7 $\pm$ 1.1 & 0.08 & 0.90 & 0.02 \\
& 0-0.2 & 90 nm & 2.7 $\pm$ 0.5 & 0.09 & 0.87 & 0.04 \\
& 0-0.5 & 210 nm & 2.2 $\pm$ 0.4 & 0.11 & 0.80 & 0.09 \\
& 1.5-2.5 & 2 $\mu$m & 1.0 $\pm$ 0.2 & 0.03 & 0.51 & 0.45 \\
&&&&&&\\
\multirow{5}{*}{Acid-cleaned synthetic (LMSY)} & 0-0.03 & 18 nm & 1.1 $\pm$ 0.2 & 0.15 & 0.84 & 0.01 \\
& 0-0.1 & 50 nm & 6.8 $\pm$ 1.4 & 0.14 & 0.85 & 0.01 \\
& 0-0.2 & 90 nm & 1.8 $\pm$ 0.7 & 0.12 & 0.82 & 0.06 \\
& 0-0.5 & 210 nm & 2.1 $\pm$ 0.4 & 0.16 & 0.70 & 0.14 \\
& 1.5-2.5 & 2 $\mu$m & 1.4 $\pm$ 0.3 & 0.04 & 0.51 & 0.45 \\
&&&&&&\\
\multirow{2}{*}{Air-oxidized synthetic (AOMSY)} & 0-0.5 & 210 nm & 1.2 $\pm$ 0.2 & 0.18 & 0.55 & 0.27 \\
& 1.5-2.5 & 2 $\mu$m & 1.0 $\pm$ 0.2 & 0.06 & 0.35 & 0.59 \\
&&&&&&\\
\multirow{2}{*}{Natural (NAT)} & 0-0.5 & 210 nm & 1.0 $\pm$ 0.2 & 0.07 & 0.89 & 0.04 \\
& 1.5-2.5 & 2 $\mu$m & 0.17 $\pm$ 0.03 & 0.10 & 0.77  & 0.13 \\

\end{tabular}
\end{ruledtabular}
\caption{X-band CW EPR results. The defect concentration values were calculated with reference to an irradiated quartz EPR standard. The spin fraction results are extracted from the 3-component EPR spin model comprised of a narrow spin-$1/2$ contribution from spins associated with defects in the diamond lattice, a broad spin-$1/2$ component associated with defects on the diamond surface and a hyperfine-coupled electron and nitrogen nucleus component associated with substitutional nitrogen atoms in the diamond lattice. For more details refer to the methods section.}
\label{TAB1}
\end{table*}

\subsection{\label{sec:level2}Dynamic nuclear polarization performance}
\begin{figure}[htb]
 \centering
   \includegraphics{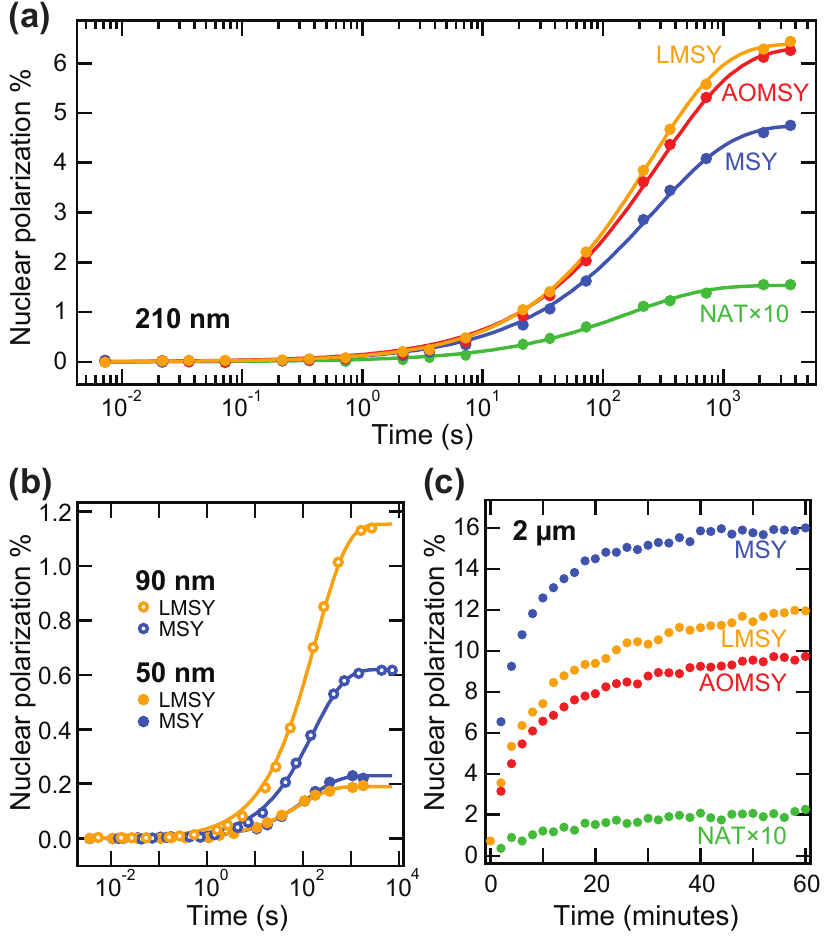}
 \caption{Dynamic nuclear polarization performance. \textbf{(a)} Saturation recovery data showing $^{13}$C polarization as a function of recovery time with microwaves driving DNP for acid-cleaned (yellow), air-oxidized (red), untreated monocrystalline synthetic (blue), and natural (green) 210 nm nanodiamond samples. \textbf{(b)} Saturation recovery DNP data for 90 nm and 50 nm acid-cleaned (yellow) and untreated monocrystalline synthetic (blue) nanodiamond samples.\textbf{(c)} Small tip angle DNP build up measurements for 2 $\mu$m diamonds. Results for the natural diamond samples are multiplied by a factor of ten to aid comparison. Solid lines show stretched exponential fits. Microwaves were tuned to the first maximum of the DNP spectrum of each sample for these measurements.}
 \label{FIG3}
\end{figure}

Having characterized the various paramagnetic defects present in our nanodiamonds we now compare those results to DNP performance. As we have shown previously, DNP performance is dependent on diamond type, size, polarizing temperature, and microwave frequency (see refs. \cite{Rej2015,Waddington2019} for more detail). Here we extend that work, showing in Fig. \ref{FIG3} and Table \ref{TAB2} how surface modification affects nanodiamond DNP. In Fig. \ref{FIG3}a we compare the DNP saturation recovery results for the same 210 nm nanodiamonds whose EPR spectra were presented in Fig. \ref{FIG2}a. Starting with the natural nanodiamond, the maximum polarization was found to be more than an order of magnitude lower than for the other diamond types. The key differences between the natural nanodiamond and its synthetic 210 nm counterparts are that it has approximately half the total number of paramagnetic defects and of those defects a smaller proportion are attributed to P1 centers. In comparison, both the synthetic samples that underwent surface treatment (increasing the proportion of P1-center spins) showed an increase in their maximum achieved polarization. Fig. \ref{FIG3}b shows a similarly improved maximum polarization for the 90 nm acid-cleaned nanodiamond sample. Again the increased maximum polarization corresponds to EPR data showing the surface treatment has narrowed the EPR lineshape and increased the fraction of spins contributed by P1 centers. These results suggest that among the mix of paramagnetic defects present in our nanodiamond samples P1 centers are more effective sites for driving DNP. The narrowing of the EPR lineshape is likely to account for a portion of the improvement in the maximum $^{13}$C polarization attained. By narrowing the EPR lineshape fewer electron spins will participate in the differential solid effect at a given DNP driving frequency. Reducing the competition between transitions that increase $^{13}$C polarization and those that  decrease $^{13}$C polarization makes the dominant polarization pathway more efficient and increases the maximum polarization that can be reached.

In contrast to the 210 nm and 90 nm samples, the nanodiamond samples with the lowest two particle sizes showed impaired DNP performance after acid cleaning. For these small nanodiamonds P1 centers contribute only 1-2\% of the electron spins and the enhancement above thermal equilibrium is very modest. Table \ref{TAB2} shows that, unlike other samples, these nanodiamonds demonstrate significantly shorter nuclear spin lattice relaxation times following surface treatment. This faster relaxation likely accounts for the bulk of the degradation in DNP performance for these samples.

Anticipating that microdiamonds would be relatively unaffected by surface treatments when compared to nanodiamonds we were surprised to find that there was a significant apparent decrease in the maximum achieved $^{13}$C polarization of the 2 $\mu$m diamond samples following surface cleaning, as shown in Fig. \ref{FIG3}c. The 2 $\mu$m diamond samples were measured using a small tip angle buildup experiment that applied a 2\degree{} pulse every two minutes during DNP buildup following an initial comb of saturation pulses. The acid-cleaned and air-oxidized microdiamonds reach a lower average $^{13}$C polarization after an hour of DNP compared to the monocrystalline synthetic microdiamond sample. The reduction in average $^{13}$C polarization however, was due to an increase in the number of NMR-visible $^{13}$C spins rather than a drop in DNP effectiveness. Surface cleaning produces an increase in the NMR signal at thermal equilibrium by a factor of approximately 50\%, whereas the DNP signal was relatively unchanged (shown in Supplemental Fig. 2 \cite{Supp}). We suggest this occurs because surface treatment increases the number of $^{13}$C nuclei outside the diffusion barrier and able to participate in the NMR signal, but that these spins are not hyperpolarized due to a combination of suppressed spin diffusion, lack of nearby P1 centers or insufficient microwave power. The natural diamond demonstrated the poorest performance of the 2 $\mu$m samples and we attribute that to a broad EPR linewidth and low concentration of P1 centers.

\begin{figure}[htb]
 \centering
   \includegraphics{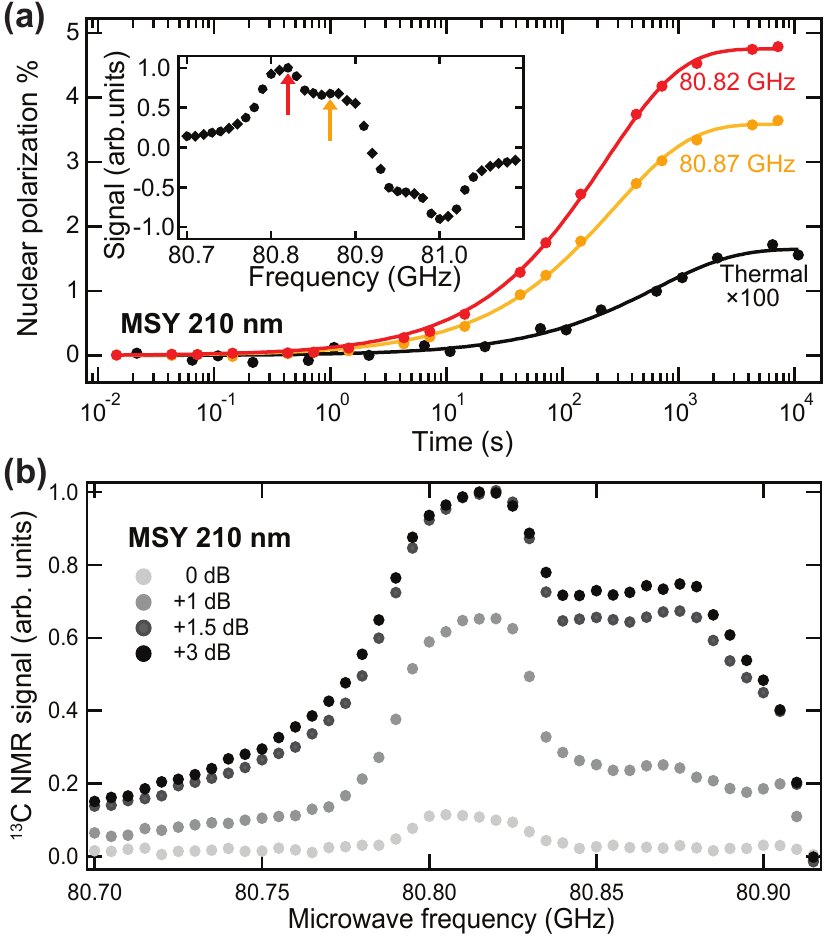}
 \caption{Frequency and power-dependent dynamic nuclear polarization behavior. \textbf{(a)} Saturation recovery data for 210 nm monocrystalline synthetic nanodiamond showing $^{13}$C polarization as a function of recovery time for DNP with microwaves at 80.82 GHz (red) and 80.87 GHz (yellow), and recovery to thermal equilibrium (black) scaled by a factor of 100. Solid lines show stretched exponential fits. Inset shows the corresponding DNP spectrum with the selected microwave frequencies indicated. \textbf{(b)} DNP spectra for 210 nm monocrystalline synthetic nanodiamond varying microwave power. As the spectra are antisymmetric only the first half of each spectrum was acquired.}
 \label{FIG4}
\end{figure}

Closely examining the DNP behavior of a single nanodiamond sample yields additional information about which paramagnetic defects provide more effective DNP pathways to high $^{13}$C polarization.   Saturation recovery experiments, carried out at the outer and inner peaks of the DNP spectrum of the 210 nm synthetic nanodiamond sample, are shown in Fig. \ref{FIG4}a and demonstrate a clear difference favoring DNP via the outer peak. Corresponding results also show that characteristic decay rates differ for nanodiamond polarized by different pathways \cite{Rej2015,Waddington2019}. We present this as evidence suggesting spin diffusion is somewhat suppressed in these nanodiamond samples, preventing bulk spins far away from paramagnetic sites from rapidly equilibrating local differences in polarization. Results comparing DNP via the inner or outer peak of the DNP spectrum for other nanodiamond samples are presented in Table \ref{TAB2}. There is a distinct trend in which higher maximum polarization is achieved via the outer peak of the DNP spectrum for those diamond samples with a sufficient concentration of P1-center defect sites. This correlation is unsurprising because the outer peaks of the DNP spectrum correspond to polarization enhancement achieved almost exclusively via electron spins associated with P1 centers. Furthermore, the outer wings of the DNP spectrum are outside the majority of the microwave frequencies at which the differential solid effect gives rise to competition with the dominant polarization pathway. With spin diffusion suppressed it is also likely that P1 centers, distributed throughout the diamond lattice, can drive DNP for a greater number of $^{13}$C spins than can paramagnetic defects near the diamond surface.

To examine the impact of varying microwave power on DNP behavior Fig. \ref{FIG4}b shows a series of DNP enhancement spectra obtained by sweeping the microwave frequency at different power levels. The spectra show that although polarization increases with microwave power across the DNP spectrum, the peak centered around 80.82 GHz shows the highest $^{13}$C signal, especially at low microwave power, and was the fastest point to reach its maximum signal. This provides further evidence that the P1 centers act as more efficient sites for DNP, producing higher $^{13}$C polarization levels with lower incident microwave power.

\subsection{\label{sec:level2}Electron relaxation}
\begin{figure}[tb]
 \centering
   \includegraphics{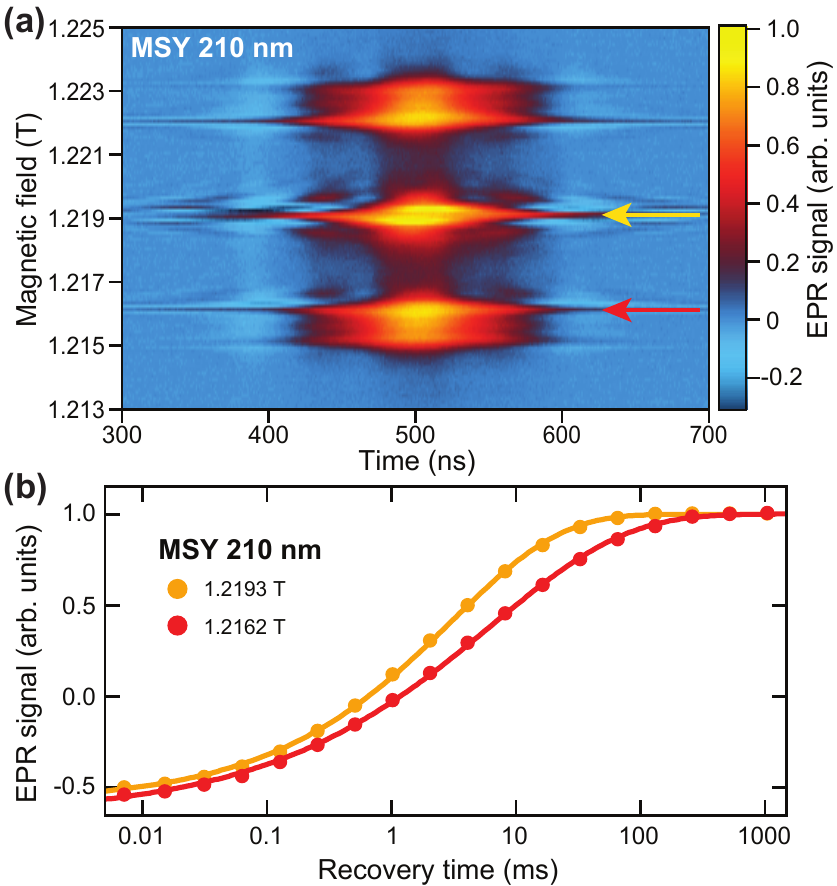}
 \caption{Pulsed electron paramagnetic resonance. \textbf{(a)} Electron spin-echo spectrum of 210 nm monocrystalline synthetic nanodiamond. Arrows indicate the central resonance (yellow) and one of the wings associated with the hyperfine splitting of the P1 center (red). \textbf{(b)} Inversion recovery results measured at the magnetic field values corresponding to the resonances marked in (a). Solid lines show stretched exponential fits.}
 \label{FIG5}
\end{figure}

To fully account for the observed variation in DNP performance between different nanodiamond samples and different defect sites we turn to our pulsed EPR results. The pulsed EPR measurements were carried out at Q-band (33.8 GHz) and 4.5 K to approach the conditions inside the DNP polarizer during operation. Pulsed EPR data for 210 nm synthetic nanodiamond is shown in Fig. \ref{FIG5} as a representative example and to correspond to the DNP results shown in Fig. \ref{FIG4}. Electron spin-echo spectra were acquired to map the X-band CW EPR data onto the Q-band results and to selectively address sections of the EPR spectrum, as shown in Fig. \ref{FIG5}a. To measure the electron spin-lattice and spin-spin relaxation times ($T_{1e}$ and $T_{2e}$ respectively) inversion recovery and spin-echo experiments were carried out with the magnetic field adjusted to bring either the central transition or one of the wings of the EPR spectrum on resonance.  In Fig. \ref{FIG5}b we show the results from two inversion recovery experiments performed at the magnetic fields marked by the arrows in Fig. \ref{FIG5}a on the same synthetic 210 nm nanodiamond sample. A quick inspection reveals that $T_{1e}$ is significantly longer for electron spins measured with the magnetic field tuned to the resonance associated with the wing of the EPR spectrum created by P1 centers. The collected fitting results, presented in Table \ref{TAB2}, show that the difference in $T_{1e}$ between the center and wing transitions is a factor of 2 or more for all samples with a reliable fit. They also show a tremendous increase in $T_{1e}$ with increasing particle size, spanning nearly 4 orders of magnitude (in comparison, there is little variation in the $T_{2e}$ results, shown in Supplemental Table. I \cite{Supp}). The $T_{1e}$ results provide compelling evidence to explain the relative DNP performance of our nanodiamond samples. In the context of DNP, electron spins with a longer $T_{1e}$ will exhibit slower diffusion of spin polarization across the EPR spectrum and their spin polarization will be more easily saturated by the microwave field used to drive DNP. Electrons with a longer $T_{1e}$ are also less effective at driving nuclear relaxation and the diffusion boundary encompassing local nuclear spins is smaller, increasing the proportion of bulk nuclear spins that can take part in DNP \cite{Wenckebach2016}. The combined effect of a longer $T_{1e}$ is an increase in the maximum achievable nuclear spin polarization via DNP. Taken together with the narrower EPR linewidth of the P1 centers relative to the nuclear Larmor frequency, and the distribution of P1 centers throughout the nanodiamond lattice, the longer $T_{1e}$ of the P1 centers provides clear evidence for the improved DNP performance of nanodiamonds with a higher proportion of P1-center spins.

\begin{table*}[htb]
\begin{ruledtabular}
\begin{tabular}{l l l l l l l}

\multirow{2}{*}{Diamond} & \multirow{2}{*}{Median Size} & \multirow{2}{*}{$^{13}$C $T_{1}$ (s)} & \multicolumn{2}{c}{Maximum $^{13}$C polarization \%} & \multicolumn{2}{c}{$T_{1e}$ ($\mu$s)}\\

 & & & Inner DNP peak & Outer DNP peak & Central resonance & P1-center wing \\
\hline
\multirow{5}{*}{MSY} & 18 nm & 60 $\pm$ 14 & 0.124 $\pm$ 0.002 & - & 60 $\pm$ 10 & - \\
& 50 nm & 200 $\pm$ 40 & 0.286 $\pm$ 0.004 & 0.230 $\pm$ 0.003 & 680 $\pm$ 20 & 2700 $\pm$ 200 \\
& 90 nm & 270 $\pm$ 80 & 0.612 $\pm$ 0.007 & 0.620 $\pm$ 0.005 & 1010 $\pm$ 20 & 2690 $\pm$ 50 \\
& 210 nm & 640 $\pm$ 100 & 3.62 $\pm$ 0.03 & 4.75 $\pm$ 0.04 & 3110 $\pm$ 70 & 6400 $\pm$ 200 \\
& 2 $\mu$m & 180 $\pm$ 30 , 3900 $\pm$ 1500 & 13.8 $\pm$ 0.2 & 16.0 $\pm$ 0.2 & 83000 $\pm$ 4000 & $>$16000 \\
&&&&&&\\
\multirow{5}{*}{LMSY} & 18 nm & 16 $\pm$ 4 & 0.037 $\pm$ 0.003 & - & 17.3 $\pm$ 0.2 & - \\
& 50 nm & 91 $\pm$ 8 & 0.179 $\pm$ 0.003 & 0.191 $\pm$ 0.002 & 220 $\pm$ 10 & 1350 $\pm$ 30 \\
& 90 nm & 370 $\pm$ 60 & 1.36 $\pm$ 0.02 & 1.15 $\pm$ 0.01 & 540 $\pm$ 10 & 2530 $\pm$ 50 \\
& 210 nm & 480 $\pm$ 180 & 3.82 $\pm$ 0.02 & 6.47 $\pm$ 0.04 & 2370 $\pm$ 60 & 5000 $\pm$ 200 \\
& 2 $\mu$m & 250 $\pm$ 60 , 4100 $\pm$ 700 & 9.8 $\pm$ 0.2 & 12.8 $\pm$ 0.2 & 76000 $\pm$ 5000 & $>$60000 \\
&&&&&&\\
\multirow{2}{*}{AOMSY} & 210 nm & 870 $\pm$ 150 & 4.79 $\pm$ 0.03 & 6.3 $\pm$ 0.04 & 7200 $\pm$ 200 & 8800 $\pm$ 400 \\
& 2 $\mu$m & 60 $\pm$ 16 , 3200 $\pm$ 400 & 9.62 $\pm$ 0.2 & 10.5 $\pm$ 0.2 & 44000 $\pm$ 4000 & $>$9000 \\
&&&&&&\\
\multirow{2}{*}{NAT} & 210 nm & 460 $\pm$ 120 & 0.15 $\pm$ 0.02 & 0.153 $\pm$ 0.002 & 1420 $\pm$ 60 & 4600 $\pm$ 300 \\
& 2 $\mu$m & $>$10000 & 0.4 $\pm$ 0.1 & 0.3 $\pm$ 0.1 & 42000 $\pm$ 4000 & $>$40000 \\

\end{tabular}
\end{ruledtabular}
\caption{Thermal $^{13}$C $T_{1}$, DNP and pulsed EPR results.  The $^{13}$C $T_{1}$ values were measured at 4.5 K, 2.89 T and quote results from stretched exponential fits to saturation recovery data or, where two values are quoted, double exponential fits. The maximum polarization percentages quoted are the values extracted from fits to polarization build up curves measured at 4.5 K, 2.89 T, such as those shown in Fig. \ref{FIG3}. The $T_{1e}$ values were measured at 4.5 K, Q-band using an echo-detected inversion recovery sequence. Results are quoted as $>$ where data could not be collected to provide fits of sufficient confidence to quote upper bounds.}
\label{TAB2}
\end{table*}

\section{\label{sec:level1}CONCLUSION AND FUTURE DIRECTIONS} 
We have tailored the composition of paramagnetic defect sites in nanodiamond and mapped how those alterations affect hyperpolarization via DNP. We have identified the importance of defect type, concentration and distribution to the effectiveness of DNP and pointed to the P1-center defect as an excellent source of electrons to drive DNP.

The logical next step is to investigate fine-tuning the concentration and distribution of P1-center defects in a high-purity nanodiamond free from significant quantities of other defect sites. The high-pressure-high-temperature diamond growth conditions that yield specific concentrations of paramagnetic defects are well understood and we anticipate that surface cleaning techniques such as those we have applied here will play an important role in eliminating undesirable surface defects \cite{Babich2000,Tsukahara2019}. The surface treatment techniques used here will also  impact approaches to developing nanodiamond for use as an imaging agent or quantum sensor that relies on the nanodiamond surface to drive hyperpolarization of nearby molecules \cite{Waddington2017,Rej2017}. 

Isotopic enrichment offers a complimentary avenue for exploration. The clear advantage of increasing the $^{13}$C concentration is a proportional increase in NMR sensitivity. Up to a limit this can be achieved without compromising the long $T_1$ relaxation time of diamond, but $T_2$ relaxation will accelerate as the dipolar coupling between $^{13}$C nuclei increases \cite{Balasubramanian2009,Shabanova1998}. The $^{13}$C concentration at which the NMR linewidth is sufficiently broadened to prohibit effective spatial encoding in MRI enforces a limit to isotopic enrichment, however, if performed in concert with fabrication of high-purity diamond this limit could be higher than 10\% \cite{Knowles2013}. A more exciting prospect for altering the DNP behavior of nanodiamond is the increase in spin diffusion anticipated following $^{13}$C enrichment \cite{Shabanova1998}. An increase in the rate of spin diffusion would accelerate the transfer of polarization away from paramagnetic sites during DNP, potentially allowing DNP to be driven efficiently with fewer defect sites and providing the benefit of reduced $T_1$ relaxation, especially at low magnetic fields \cite{Terblanche2001}. This type of behavior has been observed in silicon microparticles but in that case hyperpolarization levels suffered at the expense of extending $T_1$ \cite{Lee2011}.

We have outlined the challenge posed by the compromise between hyperpolarization and relaxation inherent to using endogenous paramagnetic defects to drive DNP. One alternative is to avoid this compromise by using exogenous or photo-induced radicals that can be removed following DNP. This form of DNP has been achieved with silicon nanoparticles using TEMPO and could be adapted for nanodiamond \cite{Hu2018}. The use of photo-induced radicals is more appealing because although the radicals are long-lived at cryogenic temperatures they spontaneously recombine at room temperature. Furthermore, the technique has shown promise in hyperpolarized $^{13}$C metabolite research \cite{Capozzi2015}. The idea could be applied to nanodiamond by leveraging the properties of the A center, a defect made up of two adjacent substitutional nitrogen atoms that is diamagnetic but becomes paramagnetic if ionized by ultraviolet light with a wavelength shorter than 415 nm \cite{Tucker1994}. 

In conclusion, we have tailored nanodiamonds to improve their performance as $^{13}$C MRI imaging agents and outlined the next steps required to create nanodiamonds capable of balancing the capacity for significant hyperpolarization with long relaxation times. We anticipate that further optimization of the spin properties of nanodiamond will lead to the creation of a useful MRI theranostic agent.

\section{\label{sec:level1}METHODS} 

\subsection{\label{sec:level2}Sample preparation}
Monocrystalline synthetic (MSY) and natural (NAT) nanodiamond powders were purchased from Microdiamant. MSY nanodiamonds were manufactured using high-pressure high-temperature synthesis. NAT nanodiamonds were produced by processing mined, industrial-quality, natural diamond. Air oxidized (AOMSY) samples were prepared by firing MSY nanodiamond in a furnace at 550\degree{}C for one hour. We estimate that the nanodiamonds decrease in size by no more than 5 nm, taking into account the furnace heating time of one hour, cool down of 20 minutes and the etch rate for diamond (4 nm/hr at 550\degree{}C, 1 nm/hr at 500\degree{}C)\cite{Gaebel2012}. Acid cleaned (LMSY) nanodiamond samples were prepared by multiple purification steps using nitric and sulfuric acid performed by Lucigem on MSY nanodiamond. Attenuated total reflectance Fourier transform infrared spectroscopy (ATR-FTIR) was performed using a Thermo Scientific Nicolet iS10 spectrometer to characterize the functional groups on the LMSY nanodiamond surface. ATR-FTIR showed an increase in the characteristic 1780 cm$^{-1}$ carbonyl peak associated with carboxylic acid groups, confirming oxidation of the nanodiamond surface \cite{Bradac2018}. 

\subsection{\label{sec:level2}CW EPR}
CW EPR spectra were measured with a Bruker EMX X-Band spectrometer operating at 9.75 GHz, and room temperature. The magnetic field modulation amplitude was set to 0.02 mT at a modulation frequency of 100 kHz. The modulation amplitude value was selected by decreasing the amplitude until further decrease caused no apparent decrease in linewidth. This avoided excessive distortion of the EPR signal whilst providing adequate signal-to-noise. Microwave power was set to 0.02 mW, which was determined to be at least 6 dB below the level at which significant saturation of the EPR spectrum could be observed in a progressive saturation study \cite{Eatons2010}. We note that the long $T_{1e}$ times we report in section II. C. explain why this low power level was required for measuring accurate EPR spectra. The sweep width was set to 20 mT, with a sweep time of 120 s and corresponding time constant of 10 $\mu$s. To quantify the concentration of electron spins, each EPR spectrum was integrated twice following a linear baseline correction. Integrated EPR signals were normalized for the sample weight and the estimated cavity Q read out from the spectrometer. An irradiated quartz EPR standard was selected as an appropriate reference for its similar electron spin properties\cite{Eaton2010}. To determine the proportional weights of different defect types contributing to the total EPR signal for each sample a three-component model was fitted to the recorded spectrum using EasySpin\cite{Stoll2006}. The three components were simulated as two spin-$1/2$ systems with a g-factor of 2.0024 and a spin system with a nitrogen nucleus introduced with a g-factor of 2.0017 and hyperfine tensor principle values A$\parallel$ =  81.5 MHz and A$\perp$ = 113.5 MHz in accordance with previous EPR studies of diamond \cite{Smith1959,Barklie1981,Fionov2010,Osipov2017}. Representative EPR fitting results corresponding to the data shown in Fig. \ref{FIG2} are shown in Supplemental Fig. 1 \cite{Supp}. The two spin-$1/2$ systems were initialized with significantly different linewidths to represent either electrons associated with defects on the nanodiamond surface (broad component) or in the nanodiamond core (narrow component). The spin system with a $^{14}$N nucleus represented P1 centers distributed throughout the nanodiamond core.

\subsection{\label{sec:level2}Pulsed EPR} 
The pulsed EPR measurements were recorded with a Bruker ELEXSYS-II E580 spectrometer operating at 33.8 GHz and 4.5 K with a Q-Band EN 5107D2 resonator in an Oxford Instruments flow cryostat. An electron spin echo spectrum was recorded for each nanodiamond sample to confirm the on-resonance magnetic field values at Q-band frequency. Calibrated $\pi/2$- and $\pi$-pulse lengths varied for each sample but for even the shortest $\pi/2$-pulses of ~20 ns the bandwidth was sufficiently narrow to selectively excite discrete ~1.8 mT sections of the EPR spectrum. The ability to selectively interrogate different sections of the EPR spectrum entails the necessary tradeoff that not all spins excited by a particular pulse will undergo the same nutation depending on their degree of detuning. This minor effect can be most readily observed in the inversion recovery data where the original $\pi$-pulse clearly does not invert all the spin polarization. To investigate the spin lifetimes of different defects the magnetic field was set to bring either the central transition or one of the wings of the EPR spectrum on resonance. To measure $T_{2e}$ a spin-echo pulse sequence was used, with the interpulse spacing increased until no echo could be detected. To exclude the defense pulse from the data at short inter-pulse spacing only the second half of the acquired echo was Fourier transformed, phased, and the peak taken as the signal. $T_{2e}$ data was fitted with a decaying exponential function to extract a characteristic decay time. For the micron-sized diamond particles a bi-exponential function provided a more appropriate fit. To measure $T_{1e}$ an echo-detected  inversion recovery sequence was implemented. The choice of an echo-detected pulse sequence was made to leverage the generous electron spin-spin relaxation time of nanodiamond to avoid the overlap of the defense pulse and the signal following the $\pi/2$-pulse. $T_{1e}$ data was fitted with a stretched exponential function. For the micron-sized diamond particles the signal did not saturate within the sequence duration limit of 1.07 s set by the spectrometer. As a result the uncertainty in $T_{1e}$ extracted from this data increased appreciably but a lower bound for $T_{1e}$ in the affected samples was still confidently established.

\subsection{\label{sec:level2}DNP}
Dynamic nuclear polarization experiments were carried out at 2.89 T, 4.5 K, in an Oxford 360/89 superconducting magnet using a scratchbuilt DNP probe, a Janis helium flow cryostat and a Tecmag Redstone spectrometer. The DNP probe incorporated a waveguide with a tunable slotted antenna to deliver 80-82 GHz microwaves to the sample in a PTFE tube located in the isothermal region of the cryostat. To transmit and receive at the $^{13}$C NMR frequency the DNP probe included a saddle coil tuned and matched to 30.9 MHz with Voltronics high-voltage trimmer capacitors. To avoid undesired coupling between the NMR coil and the microwaves the coil was oriented such that its magnetic field was perpendicular to the magnetic field axis of microwaves emitted from the slot antenna. Microwaves were generated using a Vaunix LMS163 Lab Brick signal generator driving a Virginia Diodes SGX106 amplifier multiplier chain and amplified by a Quinstar 2 W power amplifier. 

The temperature of the sample was monitored with a Lakeshore Cernox thermometer, integrated into the vaporizer assembly at the base of the cryostat, in conjunction with a Lakeshore Ruthenium oxide thermometer, mounted to the fiberglass former for the NMR coil away from the slot antenna output. The combination of these two sensors showed that under normal cryostat operating conditions with microwave irradiation on, the sample temperature was approximately 2-3 K above the temperature of the cryostat.

A DNP spectrum was recorded for each sample to determine at which frequencies maxima occurred by sweeping the microwave frequency and measuring the corresponding change in $^{13}$C signal. At each frequency step a saturation recovery pulse sequence was run with a fixed recovery time. $^{13}$C nuclear polarization was initially nulled by the application of a saturation comb of 64 $\pi/2$-pulses 10 ms apart. The polarization was then allowed to recover for 10 s before a single $\pi/2$-pulse followed immediately by acquisition. For less readily-polarized samples the recovery time was increased to provide improved signal-to-noise. The frequency was swept in 10 MHz increments over a range of 400 MHz. For sweeps performed at different microwave powers the Quinstar power amplifier was removed and the output of the Lab Brick signal generator was varied. The acquired free induction decay data (FID) was Fourier transformed and the phase adjusted so that the first maximum in the DNP spectrum was entirely in the absorption mode. The peak of the real part of the signal in frequency space was then plotted as a function of frequency to produce the DNP spectrum.

Saturation recovery scans were then performed at the maxima identified in the microwave frequency sweep data, following the saturation sequence described above, with recovery times incremented approximately logarithmically. FID data was analyzed using the same method as above and the peaks fitted with a stretched exponential. A stretched exponential function was selected to match the expected form of nuclear polarization build up driven by paramagnetic defects in solids. In such cases polarization is proportional to $e^{{-(t/\tau)}^{\alpha}})$ where $t$ is the time, $\tau$ is the time constant and $\alpha$ is a fractional power, dependent on the distribution of nuclei and paramagnetic defects, that varies between $1/2$ and $2/3$ for nanodiamond in a strong external magnetic field \cite{Blumberg1960,Furman1995,Furman1997}.

The same procedure was followed with microwaves off to record the thermal $T_1$ at 4.5 K, 2.89 T and the signal at thermal equilibrium. The Fourier transformed signal at thermal equilibrium and at maximum polarization for both frequencies investigated for each sample was then fitted with a Lorentzian fit and the comparative peak heights used to calculate the relative differences in polarization. To quantify that difference, the signal recorded at thermal equilibrium was equated with a $^{13}$C polarization of $1.65 \times 10^{-4}$ calculated from the Boltzmann distribution equation: $P = \tanh{\frac{\gamma\hbar B_0}{2k_BT}}$ where $\gamma$ is the gyromagnetic ratio of $^{13}$C, $T$ is temperature, $B_0$ is magnetic field, $\hbar$ is Planck's constant and $k_B$ is Boltzmann's constant. This allowed all saturation recovery scans to be plotted as a function of $^{13}$C polarization, providing a clear and objective measure of DNP performance. A similar procedure was followed for the 2 $\mu$m diamond powders with the exception that DNP build up was measured with a small tip angle sequence consisting of a 2$^{\circ}$ tip every 2 minutes calibrated using a single 20-minute build up from saturation.

\begin{acknowledgements}
The authors would like to acknowledge the support of the Australian Research Council Centre of Excellence Scheme (Grant No. EQuS CE110001013) ARC DP1094439. We also acknowledge use of the EPR tools provided by the NMR Facility within the Mark Wainwright Analytical Centre at the University of New South Wales.
\end{acknowledgements}

\bibliography{Bibliography} 

\end{document}